\documentstyle[aps,prd]{revtex}
\renewcommand{\narrowtext}{}
\renewcommand{\mediumtext}{}
\begin{document}

\title{On the Particle Data Group evaluation of  
\mbox{$\chi_c$} and 
\mbox{$\psi'$}
branching ratios}

\author{Claudia Patrignani} 
\address{Universit\`a\ di Genova and I.N.F.N. Sezione di Genova\\
Via Dodecaneso 33, I-16146 Genova, Italy\\
E-mail: patrignani@ge.infn.it}

\date{\today}
\maketitle
\begin{abstract}
I propose a new evaluation of  
$\psi'$ and $\chi_c$ branching ratios which avoids the correlations
affecting the current Particle Data Group evaluation.

These correlations explain the apparent 
technique-dependent discrepancies between the available determinations 
of the ${\cal B}(\chi_c\to p\bar p)$ and $\Gamma(\chi_c\to \gamma\gamma)$
under the hypothesis that the current values of
the $\psi'\to\chi_c\gamma$ branching ratios are overestimated.
 
In the process I also noticed that
the Particle  Data Group  has not  restated many of the older
measurements, when necessary, for the new value of ${\cal B}(J/\psi\to
l^+l^-)$, which significantly affects the evaluation of some relevant $\psi'$ and 
$\chi_c$ exclusive branching ratios.

\end{abstract}
\pacs{13.20.Gd,14.40.Gx}
\narrowtext
\section{INTRODUCTION}
All of the older measurements of $\chi_c$ branching ratios where
performed in $e^+e^-$ experiments, where the $\chi_c$ were studied
in the $\psi'$ radiative decays.

In recent years new measurements appeared for ${\cal B}(\chi_c\to
p\bar p)$ and $\Gamma(\chi_c\to \gamma\gamma)$ obtained in $e^+e^-$, $\bar p p$
and  $e^+e^-\to e^+e^-\gamma\gamma$ experiments.

Although the accuracy of these measurements is limited, it's becoming
evident that the values of ${\cal B}(\chi_c\to p\bar p)$ derived by measurements
performed in
$\bar p p$ annihilation experiments tend to be systematically higher
than the values obtained in $e^+e^-\to\psi'$  experiments
(Tab.\ \ref{tab:compp}), while for the $\Gamma(\chi_c\to \gamma\gamma)$
the values derived by measurements in $\bar p p$ experiments are significantly lower
than the values obtained in $e^+e^-\to e^+e^-\gamma\gamma$ experiments
(Tab.\ \ref{tab:compg}).

I reviewed the procedure and the data
used by the Particle Data Group (PDG\cite{bib:PDG00}) to evaluate these
quantities and I realized that this peculiar pattern of 
technique-dependent discrepancies could be
explained by a systematic error in 
${\cal B}(\psi'\to \chi_c\gamma)$.

The problem is however of a more general kind, since it depends on the
fact that in many cases the PDG uses in its evaluation quantities
that are derived from the measurement. At least for some of the $\psi'$
and $\chi_c$ branching ratios,
this procedure introduces a correlation among the values obtained by
different experiments which has not been properly taken into account.

I also noticed 
that in most cases the older data used by 
PDG 
to evaluate the $\psi'$ branching ratios to final states including a
$J/\psi$ and the $\chi_c$ radiative decay branching ratios had not
been restated for the current world average of ${\cal B}(J/\psi\to
l^+l^-)$, whose current central
value is now about 15\% lower than the value currently accepted at
the time most of the measurements were performed.

While this only involves a restatement of a number of experimental results, 
it affects the value of exclusive $\psi'$ and $\chi_c$ branching 
ratios which are relevant for a large number of other experiments.

The outline of the paper is as follows:
in section \ref{sec:ratio} the problems found in the
evaluation of $\psi'$ and $\chi_c$ branching ratios are discussed;
in section \ref{sec:chibr} a new evaluation is presented.
The results are discussed in section \ref{sec:concl}  and compared with some 
theoretical expectations for the partial widths of the radiative transitions.

\section{RATIONALE FOR A NEW EVALUATION OF
${\lowercase{\psi}}'$,
${\lowercase{\chi_c}}$ BRANCHING RATIOS} 
\label{sec:ratio}
There is often an ambiguity in presenting the experimental result between
the measurement and the determination of what the experiment considers the
relevant branching ratio, extracted from its measurement based on other, 
better known quantities, to which the result is clearly correlated. 

While this serves the legitimate purpose of showing the impact of a new 
measurement, it might contribute to an improper treatement of data 
by PDG if the correlation to other measurements is not
given enough evidence.

As an example, the PDG compilation lists as independent measurement of 
${\cal B}(\chi_{c_0}\to p\bar p)$, 
the values quoted by BES\cite{bib:Bai98I} and E835\cite{bib:Ambrogiani99}.

E835 explicitly mentions in the abstract that the value they quote for  
${\cal B}(\chi_{c_0}\to p\bar p)$ is derived from a direct measurement of
\begin{equation}
\Gamma(\chi_{c_0}\to p\bar p){\cal B}(\chi_{c_0}\to J/\psi\gamma)
{\cal B}(J/\psi\to e^+e^-)=2.89^{+0.67}_{-0.53}\pm0.14\,eV\label{eq:Ambrogiani99},
\end{equation}
while it is not immediate to recognize that the value of 
${\cal B}(\chi_{c_0}\to p\bar p)$ quoted by BES is obtained by a measurement of
\begin{equation}
\frac{{\cal B}(\psi'\to \chi_{c_0}\gamma){\cal B}(\chi_{c_0}\to p\bar p)}
{{\cal B}(\psi'\to J/\psi \pi^+\pi^-)} =
(4.6\pm1.9)\times 10^{-5}\label{eq:Bai98}
\end{equation}
because this value is not explicitly quoted in \cite{bib:Bai98I}
and the value in eq. (\ref{eq:Bai98}) must be calculated from
the number of $\psi' \to \chi_{c_0} \gamma \to p \bar p  \gamma$ 
decays observed, the corresponding efficiency, and the total number of $\psi'$, which
in turn, as specified in one of their references \cite{bib:Bai98D}, is obtained from  
the observed number of $\psi' \to J/\psi \pi^+\pi^-$ using for this decay mode
the value of branching ratio taken from the 1996 compilation of PDG 
(PDG96)\cite{bib:PDG96}. 

The value of ${\cal B}(\chi_{c_0}\to p\bar p)$ derived from eq. 
(\ref{eq:Ambrogiani99}) is strongly anticorrelated to that obtained
from eq. (\ref{eq:Bai98}), since in the 
first case the measurement is multiplied by $\Gamma(\chi_{c_0}\to J/\psi\gamma)^{-1}$
which is derived as 
\begin{equation}
\Gamma(\chi_{c_0}\to J/\psi\gamma)^{-1}=\frac{{\cal B}(\psi'\to\chi_{c_0}\gamma)}
{{\cal B}(\psi'\to\chi_{c_0}\gamma\to J/\psi\gamma\gamma)}
\frac{1.}{\Gamma(\chi_{c_0})},
\label{eq:fAmbrogiani99}
\end{equation} 
while in the second case it is multiplied by
\begin{equation}
\frac{1.}
{{\cal B}(\psi'\to\chi_{c_0}\gamma)}\label{eq:fBai98}.
\end{equation}

The multiplicative
factor is directly proportional to ${\cal B}(\psi'\to\chi_c\gamma)$
for $p\bar p$ annihilation experiments and
inversely proportional to it for $e^+e^-\to\psi'$ experiments.
(The same applies of course to the various determinations of 
${\cal B}(\chi_{c_{1,2}}\to p\bar p)$).

Even if it is not immediately evident,
the same kind of technique-dependent anticorrelation is found among the   
different $\Gamma(\chi_{c_{0,2}}\to\gamma\gamma)$ determinations, 
as discussed in more detail in the appendix. The multiplicative
factor in this case is directly proportional to ${\cal B}(\psi'\to\chi_c\gamma)$
for $e^+e^-\gamma\gamma$ experiments and inversely proportional to it
for $p\bar p$ annihilation experiments.

The most interesting consequence of this correlation is that a systematic error
in ${\cal B}(\psi'\to\chi_c\gamma)$ would induce a peculiar pattern
of apparent technique-dependent discrepancies in both 
${\cal B}(\chi_{c}\to p\bar p)$ and $\Gamma(\chi_{c}\to\gamma\gamma)$ obtained
by this procedure.

The observed pattern shown in tabs.\ \ref{tab:compp} and \ref{tab:compg} suggests
the hypothesis
that the current values of  ${\cal B}(\psi'\to \chi_c\gamma)$ are
overestimated.

This is not the only problem in the evaluation of $\psi$ and $\chi_c$ 
branching ratios.

An inconsistency was already noticed by Gu and Li\cite{bib:Gu99} in
the 1998 edition of the PDG compilation (PDG98)\cite{bib:PDG98} concerning the 
evaluation of some $\psi'$ branching ratios which included the
measurements performed by E760\cite{bib:Armstrong97}.

PDG98, based on the 
results quoted in the abstract of the paper, listed the E760 result as 
independent measurements of 
${\cal B}(\psi'\to J/\psi\pi^+\pi^-)$, ${\cal B}(\psi'\to J/\psi\pi^0,\pi^0)$, 
and ${\cal B}(\psi'\to J/\psi\eta)$, 
while the experiment, as specified in the paper, indeed measured 
the ratio of these branching ratios to ${\cal B}(\psi'\to J/\psi\, X)$,
whose values are not explicitly quoted in the paper and must be calculated
from efficiency and event ratios.

The inconsistency in this case was due to a circular dependence, since
the multiplicative factor used in PDG98
to derive the above branching ratios from the E760 measurement, 
${\cal B}(\psi'\to J/\psi\, X)$, was itself correlated to the E760 measurement,
since it was obtained as
\widetext
\begin{eqnarray}
{\cal B}(\psi'\to J/\psi\, X)&=& {\cal B}(\psi'\to
J/\psi\pi^+\pi^-)+{\cal B}(\psi'\to J/\psi\pi^0\pi^0)  + {\cal
B}(\psi'\to J/\psi\eta) \nonumber \\ &&+{\cal
B}(\psi'\to\chi_{c_1}\gamma){\cal B}(\chi_{c_1} \to J/\psi\gamma)+ {\cal
B}(\psi'\to\chi_{c_2}\gamma){\cal B}(\chi_{c_2} \to J/\psi\gamma)
\label{eq:new}
\end{eqnarray}
\narrowtext
based on values for the
${\cal B}(\psi'\to J/\psi\pi^+\pi^-)$, 
${\cal B}(\psi'\to J/\psi\pi^0\pi^0)$, and ${\cal B}(\psi'\to J/\psi\eta)$
which included the E760 result in the average.

The problem here is rather subtle: the  values of branching ratios quoted in 
the abstract by E760 are correlated to
previous measurements, but there is no circular dependence since 
the value of ${\cal B}(\psi'\to J/\psi\, X)$, taken from PDG96\cite{bib:PDG96},
is not correlated to the E760 measurement.
The same procedure is also used
in a recent E835 paper\cite{bib:Ambrogiani00} 
reporting new, more precise 
measurements of 
$\frac{{\cal B}(\psi'\to e^+e^-)}{{\cal B}(\psi'\to J/\psi\, X)}$,
$\frac{{\cal B}(\psi'\to J/\psi\pi^0\pi^0)}{{\cal B}(\psi'\to J/\psi\, X)}$,
and $\frac{{\cal B}(\psi'\to J/\psi\eta)}{{\cal B}(\psi'\to J/\psi\, X)}$:
the quantities 
directly measured are clearly specified, but 
the abstract quotes only the non-independent estimate of
branching ratios
derived using the value of ${\cal B}(\psi'\to J/\psi\, X)$ 
from PDG96. 

Following Gu and Li,
the 2000 edition of the PDG compilation uses in its evaluation the E760 measurement of
$\frac{{\cal B}(\psi'\to J/\psi\eta)}{{\cal B}(\psi'\to J/\psi\, X)}$
and the value of 
$\frac{{\cal B}(\psi'\to J/\psi\pi^0\pi^0)}{{\cal B}(\psi'\to J/\psi\pi^+\pi^-)}$
derived from the E760 measurement, because they claim this makes the E760 input data 
independent. While this approach avoids the inconsistency described above, 
it doesn't make proper use of measurements that constraint which fractions of
${\cal B}(\psi'\to J/\psi\, X)$ are due to 
${\cal B}(\psi'\to J/\psi\pi^0\pi^0)$ and ${\cal B}(\psi'\to J/\psi\pi^+\pi^-)$
separately, not only their ratio.

An analogous problem of circular dependence is found in the 2000 edition 
of PDG in the ${\cal B}(\psi'\to J/\psi\eta)$ and
${\cal B}(\chi_{c_{1,2}}\to J/\psi\gamma)$ evaluations.

The MARK-II results on these branching ratios\cite{bib:Himel80} depend
on the value of ${\cal B}(\psi'\to J/\psi \pi^+\pi^-)$ (this decay mode is
used to determine the total number of $\psi'$ as specified in their ref. [10]),
whose average value is no longer
independent (because of the E760 measurement) from the average value of
${\cal B}(\psi'\to J/\psi\, X)$, which in turn depends on the average value
${\cal B}(\psi'\to J/\psi\eta)$ and
${\cal B}(\chi_{c_{1,2}}\to J/\psi\gamma)$, that include the MARK-II 
result.

Notice that in the last two cases the circular dependence arises from
the choice of using the value ${\cal B}(\psi'\to J/\psi\, X)$ 
determined by eq.\ (\ref{eq:new}) rather than the average of 
direct measurements. 

The correlation between the determined values of  
${\cal B}(\psi'\to J/\psi \pi^+\pi^-)$, ${\cal B}(\psi'\to J/\psi \pi^0\pi^0)$, 
${\cal B}(\psi'\to J/\psi \eta)$, and  ${\cal B}(\chi_{c_{1,2}}\to J/\psi\gamma)$
is unavoidable given that they are derived from measurements of their ratios
to ${\cal B}(\psi'\to J/\psi\, X)$ and/or ${\cal B}(\psi'\to J/\psi \pi^+\pi^-)$.

On the other side, the PDG has excluded from its averages the value of
${\cal B}(\psi'\to e^+e^-)$ derived from the E760 measurement of
$\frac{{\cal B}(\psi'\to e^+e^-)}{{\cal B}(\psi'\to J/\psi\, X)}$ while in
this case the determined value, although dependent on
${\cal B}(\psi'\to J/\psi\, X)$, is not correlated to the (less precise) 
value of ${\cal B}(\psi'\to e^+e^-)$ derived from the line shape analysis.

For all of these reasons, and also because the PDG missed to restate quite a 
few older measurements for the new value of ${\cal B}(J/\psi\to l^+l^-)$, the
$\psi'$ and $\chi_c$ branching ratios must be re-evaluated.

In order not to incurr in the kind of inconsistencies described above, 
the most convenient approach is to
use only the products of branching ratios directly measured by the experiments.

Unfortunately this implies that it is no longer possible to evaluate the $\chi_c$ 
branching ratios independently from the $\psi'$ branching ratios.

\section{A NEW EVALUATION OF $\lowercase{\psi'}$,
$\lowercase{\chi_c}$ BRANCHING RATIOS} 
\label{sec:chibr}

A fit to the $\psi'$ and $\chi_c$ branching ratios simultaneously based on all
the measurements listed in Tabs.\ \ref{tab:chi0}--\ref{tab:chi2} and 
\ref{tab:measp}\footnote{The new E835 measurements on 
$\psi'$\protect\cite{bib:Ambrogiani00} and 
$\chi_c\to\gamma\gamma$\protect\cite{bib:Ambrogiani00B} are included. The last table includes 
also some measurements already listed in Tabs.\,\ref{tab:chi1}--\ref{tab:chi2}.} 
is performed. 

In these tables I've listed the quantities directly measured by each 
experiment, even if the results are usually presented using the 
procedure discussed above. 

When the quantity directly measured is not explicitly quoted in the paper, 
I calculated it either from the
number of events and efficiencies or by rescaling the
final result by the 
values of the branching ratios used in its derivation, as specified in
the tables' footnotes.

The values have been corrected when appropriate for the current world average of 
${\cal B}(J/\psi\to l^+l^-)$.

The fit is done by a standard $\chi^2$ minimization using the MINUIT program
\cite{bib:minuit}. The $\chi^2$ is defined as:
\begin{equation}
\chi^2=\sum_i \frac{(M_i-F_i(\vec\theta))^2}{\sigma^2_i}
\end{equation}
where $M_i$ are the quantities directly
measured by each experiment, and
$F_i$ are the corresponding expected values given the set of parameters 
$\vec\theta$
\widetext
\[
\vec\theta=
\Bigl({\cal B}(\psi'\to J/\psi\pi\pi),{\cal B}(\psi'\to J/\psi\eta),
{\cal B}(\psi'\to\chi_c\gamma), {\cal B}(\chi_c\to J/\psi\gamma), 
{\cal B}(\chi_c\to p\bar p),
{\cal B}(\chi_c\to\gamma\gamma),
\Gamma(\chi_c) 
\Bigr). 
\]
\narrowtext
In all cases systematic and statistical errors
have been added in quadrature. No attempt has been made to take into account the
correlation between systematic errors on 
measurements performed by the same experiment.

The fit uses 63 measurements to fix 17 free parameters and the results
are listed in Tab.\ \ref{tab:fglob}. It has a $\chi^2$=56.0 for 46 degrees
of freedom which corresponds to  a probability of 14.8\% under the assumption of
gaussian errors. 

Given the large number of free parameters, I verified the stability of the results 
by fixing some of them to their current PDG averages.

I've chosen the variables to fix as those for which an average based on
direct independent measurements is available.

Fixing the values of the total widths $\Gamma(\chi_c)$ 
the number of free parameters is reduced to 14. 

I've also fixed the values of the ${\cal B}(\psi'\to\chi_c\gamma)$'s
to their current PDG averages. 

The value of 
${\cal B}(\psi'\to J/\psi\pi^+\pi^-)$ is weakly correlated to  
${\cal B}(\psi'\to\chi_{c_{1,2}}\gamma\to J/\psi\gamma\gamma)$ through
the ${\cal B}(\psi'\to J/\psi\, X)$  
which is calculated by eq.\ (\ref{eq:new}).

Ignoring this weak correlation and 
fixing $\langle{\cal B}(\psi'\to J/\psi\pi^+\pi^-)\rangle$
to the central value of the only direct measurement, 
performed by MARK-I back in 1975\cite{bib:Abrams75B}, 
the complexity of the fit is greatly reduced.

For this test a separate fit to each one of
the $\chi_c$ states can be performed using only the measurements listed in 
Tabs.\ \ref{tab:chi0}--\ref{tab:chi2}. Each fit has now at
at most 5 free parameters:
\widetext
\[
\vec\theta'=
\Bigl(
{\cal B}(\psi'\to\chi_c\gamma), {\cal B}(\chi_c\to J/\psi\gamma), 
{\cal B}(\chi_c\to p\bar p),
{\cal B}(\chi_c\to\gamma\gamma),
\Gamma(\chi_c) 
\Bigr). 
\]
\narrowtext

In all cases the results are stable, with the only exception of the fit in which the 
$\psi'$
radiative decay branching ratios are fixed to their current world average. 
The $\chi^2$ is also worse in this case, as would be expected in case of a systematic error
on these quantities.

\section{DISCUSSION OF THE RESULTS}
\label{sec:concl}
As a result of the fit I obtain values for the branching ratios of the radiative $\psi'$ and
$\chi_c$ decays that are significantly different from the current PDG average.

The difference between the $\chi_c$ radiative branching ratios
$(1.2^{+0.3}_{-0.2})\%$ ($\chi_{c_0}$),        
$(31.8^{+3.6}_{-3.1})\%$ ($\chi_{c_1}$), and   
$(18.7^{+2.8}_{-2.3})\%$ ($\chi_{c_2}$)        
and the corresponding current PDG averages
$(0.66\pm0.18\pm0.07)\%$,                      
$(27.3\pm1.6\pm2.5)\%$ and                     
$(13.5\pm1.1\pm1.4)\%$                         
can only in part be ascribed to the new value of ${\cal B}(J/\psi\to l^+l^-)$, 
which only accounts for an increase by 15\%.

The differences between the
new determination of the $\psi'$ radiative branching ratios,
$(7.1\pm1.2)\%$,       
$(8.4\pm0.8)\%$ and    
$(6.8\pm0.8)\%$ for    
$\chi_{c_0}$, $\chi_{c_1}$  and $\chi_{c_2}$ respectively, and the corresponding
PDG averages of 
$(9.3\pm0.9)\%$,       
$(8.7\pm0.8)\%$ and    
$(7.8\pm0.8)\%$        
can only be explained in term of a systematic error in the data, since 
in this case the PDG performs an average of genuinely independent, and direct 
measurements of these quantities.

It is interesting to notice that for all of the above branching ratios the precision
on the fit result is worse than that quoted by PDG, with the only exception of the
$\chi_{c_0}\to J/\psi\gamma$. This could be a symptom of discrepancies in the data.

The above results for the $\psi'$ branching ratios can also be interpreted as 
a confirmation that for $\chi_{c_0}$ and $\chi_{c_2}$ the technique-dependent 
discrepancies between the 
$\Gamma(\chi_{c_2}\to\gamma\gamma)$ and 
${\cal B}(\chi_{c_0}\to p\bar p)$  determinations are significant enough
to support the hypothesis that the current values
of ${\cal B}(\psi'\to\chi_{c_{0,2}}\gamma)$ (whose average is dominated by the
Crystall Ball measurement \cite{bib:Gaiser86}) are overestimated by 
$\sim$10--20\%.

The different measurements involving the ${\cal B}(\chi_{c_1}\to p\bar p)$, 
given the
present accuracy, are not sensible to an hypotetical systematic error 
in ${\cal B}(\psi'\to\chi_{c_1}\gamma)$. 

The hypothesis of a systematic error on the ${\cal B}(\psi'\to\chi_{c}\gamma)$
would also explain why the error on the
${\cal B}(\chi_c\to p\bar p)$ and 
${\cal B}(\chi_c\to \gamma\gamma)$ is sometimes drastically reduced.

For the $p\bar p$ branching ratios I obtain for $\chi_{c_0}$, $\chi_{c_1}$, and $\chi_{c_2}$
$(2.4\pm0.6)\times 10^{-4}$,
$(7.1^{+1.5}_{-1.2})\times 10^{-5}$ and
$(7.4^{+1.4}_{-1.2})\times 10^{-5}$  respectively.

These values must be compared with 
the corresponding current PDG evaluation:
$(2.2\pm1.3\pm0.2)\times 10^{-4}$,
$(8.2\pm1.3\pm0.8)\times 10^{-5}$ and
$(9.8\pm1.0\pm1.0)\times 10^{-5}$. 
For the $\chi_{c_{0}}$ the error is reduced from $\sim$60\% to $\sim$25\%.

For the ${\cal B}(\chi_c\to \gamma\gamma)$ the increased precision in the new determination,
$(2.0^{+1.0}_{-0.9})\times10^{-4}$ and
$(2.2\pm0.4)\times10^{-4}$ (for $\chi_{c_0}$ and $\chi_{c_2}$ respectively), 
with respect to the corresponding PDG evaluation 
$(2.7\pm1.9\pm0.3)\times10^{-4}$ and
$(1.6\pm0.5\pm0.2)\times10^{-4}$ is only in part due to the new
E835 measurements\cite{bib:Ambrogiani00B}.

The reasonable agreement between predicted and measured partial widths,
in particular for the radiative decay modes, has always been regarded as
one of the most remarkable features of the quarkonium model.

The values obtained here for the $\psi'$ and $\chi_c$ radiative
decays  are in
general reasonably compatible with theoretical predictions of the radiative
widths for these states as shown in Tab.\ \ref{tab:Theo}, at least to the
same level as the current PDG values.

\section{Conclusions}
I have presented a new evaluation of some $\psi'$ and $\chi_c$ branching ratios
which is, I believe, more adequate to evaluate them based
on a set of independent measurements which involve different combinations of $\psi'$ and
$\chi_c$ branching ratios simultaneously.

The new values for the
${\cal B}(\psi'\to \chi_{c_{0,2}}\gamma)$'s and ${\cal B}(\chi_{c_{0,2}}\to J/\psi\gamma)$'s
obtained here are significantly different from the current PDG 
averages.

It must be noticed that the current world average of the
$\psi'$ and $\chi_c$ radiative decays are all dominated by the
Crystall Ball measurement\cite{bib:Gaiser86}\cite{bib:Oreglia82}, and that if there is
indeed a systematic error, it is not unreasonable to expect that it be common 
to all $\psi'\to\chi_c\gamma$ and $\chi_c\to J/\psi\gamma$ measurements. 

New, more precise measurements of 
${\cal B}(\psi'\to\chi_{c_1}\gamma\to p\bar p\gamma)$ could help
to clarify the problem.

Also new measurements 
in $p\bar p$ or $\gamma\gamma$ experiments of 
selected $\chi_c$ branching ratios to decay modes other than
$J/\psi\gamma$ or $\gamma\gamma$ could provide valuable information.
  
As a last remark,
I must mention that it has not always been easy to extract
the values directly measured from the original papers, since they are
sometimes not explicitly quoted, and in a few cases relevant information
had to be gathered from references.

As shown by the case presented here, the ambiguity between measurement and
values derived from it may hide the correlations, which must be understood 
to properly analyze the data from different experiments.

\section*{Acknowledgements}
I wish to thank my colleagues from experiment E835 at Fermilab for stimulating 
discussions and comments. 
\appendix
\section*{THE PDG EVALUATION OF ${\cal B}(\lowercase{\chi_c\to p\bar p})$ AND 
$\Gamma(\lowercase{\chi_c}\to\gamma\gamma)$ }
\label{sec:ggpp}
 
The evaluation of $\langle{\cal B}(\lowercase{\chi_c\to p\bar p})\rangle$ and 
$\langle\Gamma(\lowercase{\chi_c}\to\gamma\gamma)\rangle$ is based on measurement
performed with a variety of techniques.

The values of ${\cal B}(\chi_{c_{0,1,2}}\to p\bar p)$ quoted in the original
papers 
are extracted from the measurements according to
\widetext
\begin{equation}
{\cal B}(\chi_{c}\to p\bar p)=\left\{
\begin{array}{l c l}
{\cal B}(\psi'\to \chi_c\gamma) {\cal B}(\chi_c\to p\bar p) 
   \displaystyle\frac{1.}{\langle{\cal B}(\psi'\to \chi_c\gamma)\rangle} & &(e^+e^-) \\
\\
\Gamma(\chi_c\to p\bar p) {\cal B}(\chi_c\to J/\psi\gamma)
   \displaystyle\frac{1.}{\langle\Gamma(\chi_c)\rangle \langle{\cal B}(\chi_c\to J/\psi\gamma)\rangle} & ~ & (\bar p p)
\end{array}
\right.
\label{eq:Bpp}
\end{equation}
where $\langle\rangle$ indicates the world average for the given
quantity.

Analogously the values quoted for $\Gamma(\chi_{c_{0,2}}\to \gamma\gamma)$ are  obtained 
as
\begin{equation}
\Gamma(\chi_{c}\to \gamma\gamma)=\left\{
\begin{array}{l c l}
{\cal B}(\psi'\to\chi_c\gamma) {\cal B}(\chi_c\to \gamma\gamma)
     \displaystyle\frac{\langle\Gamma{\chi_c}\rangle}{\langle{\cal B}(\psi'\to\chi_c\gamma)\rangle} & ~ & (e^+e^-) \\
\\
\Gamma(\chi_{c}\to \gamma\gamma){\cal B}(\chi_c\to J/\psi\gamma) 
     \displaystyle\frac{1.}{\langle{\cal B}(\chi_c\to J/\psi\gamma)\rangle} && (\gamma\gamma) \\
\\
{\cal B}(\chi_c\to p\bar p){\cal B}(\chi_c\to \gamma\gamma)
     \displaystyle\frac{\langle\Gamma{\chi_c}\rangle}{\langle{\cal B}(\chi_c\to p\bar p)\rangle} && (\bar p p) \\
\\
\displaystyle\frac{{\cal B}(\chi_c\to \gamma\gamma)}{{\cal B}(\chi_c\to J/\psi\gamma)}
  \langle{\cal B}(\chi_c\to J/\psi\gamma)\rangle\langle\Gamma(\chi_c)\rangle && (\bar p p)
\end{array}
\right.
\label{eq:Ggg}
\end{equation}
\narrowtext

The PDG evaluates the $\langle{\cal B}(\chi_{c_{0,1,2}}\to p\bar p)\rangle$ 
and $\langle\Gamma(\chi_{c_{0,2}}\to \gamma\gamma)\rangle$  
averaging the values derived by the different experiments according to
eqs.\ (\ref{eq:Bpp}) and (\ref{eq:Ggg})\footnote{The last expression in Eq.\ (\ref{eq:Ggg}) is used in a recent E835
paper\cite{bib:Ambrogiani00B} to present a new determination of
$\Gamma(\chi_{c_{0,2}}\to \gamma\gamma)$. 
It is not yet included in PDG averages.}.

There is more than one reason to object to this procedure.

First of all the
error attributed to the $\langle{\cal B}(\chi_c\to J/\psi\gamma)\rangle$, 
calculated by 
\begin{equation}
{\cal B}(\chi_{c}\to J/\psi\gamma)=
\frac{{\cal B}(\psi'\to\chi_{c}\gamma\to\chi_{c}\to J/\psi\gamma\gamma)}
{{\cal B}(\psi'\to\chi_{c}\gamma)}\label{eq:bchi}
\end{equation}
doesn't even include the $\sim$10\% uncertainty in $\langle{\cal B}(\psi'\to \chi_c\gamma)\rangle$ which 
is the dominant source of error on the $\langle{\cal B}(\chi_c\to J/\psi\gamma)\rangle$.

A more serious objection regards the fact that the different values of 
${\cal B}(\chi_c\to  p\bar p)$ and 
$\Gamma(\chi_c\to \gamma\gamma)$ obtained in this way are not 
independent.

If we substitute Eq.\ (\ref{eq:bchi}) in Eqs.\ (\ref{eq:Bpp}) and
(\ref{eq:Ggg}), 
and consider that the $\langle{\cal B}(\chi_c\to  p\bar p)\rangle$ is dominated by the
measurements performed in $\bar p p$ experiments, we obtain
\widetext
\begin{equation}
{\cal B}(\chi_{c}\to p\bar p)=\left\{
\begin{array}{l c l}
{\cal B}(\psi'\to \chi_c\gamma) {\cal B}(\chi_c\to p\bar p) 
   \displaystyle\frac{1.}{\langle{\cal B}(\psi'\to \chi_c\gamma)\rangle} && (e^+e^-)\\
\\
\Gamma(\chi_c\to p\bar p){\cal B}(\chi_c\to J/\psi\gamma)
  \displaystyle\frac{1}{\langle\Gamma(\chi_c)\rangle}\frac{\langle{\cal B}(\psi'\to\chi_c\gamma)\rangle}
  {\langle{\cal B}(\psi'\to \chi_c\gamma\to J/\psi\gamma\gamma)\rangle} & ~ & (\bar p p) 
\end{array}
\right.
\label{eq:BNpp}
\end{equation}

\begin{equation}
\Gamma(\chi_{c}\to \gamma\gamma)=\left\{
\begin{array}{l c l}
{\cal B}(\psi'\to\chi_c\gamma) {\cal B}(\chi_c\to \gamma\gamma)
     \displaystyle\frac{\langle\Gamma{\chi_c}\rangle}{\langle{\cal B}(\psi'\to\chi_c\gamma)\rangle} & ~ & (e^+e^-) \\
\\
\Gamma(\chi_{c}\to \gamma\gamma){\cal B}(\chi_c\to J/\psi\gamma)
  \displaystyle\frac{\langle{\cal B}(\psi'\to \chi_c\gamma)\rangle}
  {\langle{\cal B}(\psi'\to \chi_c\gamma\to J/\psi\gamma\gamma)\rangle} && (\gamma\gamma) \\
\\
{\cal B}(\chi_c\to \gamma\gamma){\cal B}(\chi_c\to p\bar p)
  \displaystyle\frac{\langle\Gamma(\chi_{c})\rangle^2}{
   \langle\Gamma(\chi_c\to p\bar p){\cal B}(\chi_c\to J/\psi\gamma)\rangle}
   \frac{\langle{\cal B}(\psi'\to \chi_c\gamma\to J/\psi\gamma\gamma)\rangle}
  {\langle{\cal B}(\psi'\to \chi_c\gamma)\rangle}
&& (\bar p p) \\
\\
\displaystyle\frac{{\cal B}(\chi_c\to \gamma\gamma)}{{\cal B}(\chi_c\to J/\psi\gamma)}
\langle\Gamma(\chi_c)\rangle
  \displaystyle\frac{\langle{\cal B}(\psi'\to \chi_c\gamma\to J/\psi\gamma\gamma)\rangle}
  {\langle{\cal B}(\psi'\to \chi_c\gamma)\rangle}&& (\bar p p)
\end{array}
\right.
\label{eq:GNgg}
\end{equation}
\narrowtext

Not only we have a correlation between ${\cal B}(\chi_{c_2}\to  p\bar p)$ and 
$\Gamma(\chi_{c_2}\to \gamma\gamma)$, but all of the values 
are correlated  to $\langle{\cal B}(\psi'\to \chi_c\gamma)\rangle$, directly or 
through the $\langle{\cal B}(\chi_c\to J/\psi\gamma)\rangle$.

It must be noticed that this correlation has been ignored already in the 
original literature, where the values of branching ratios
extracted from the measurements according to eqs.\ (\ref{eq:Bpp}) and (\ref{eq:Ggg}) are
sometimes compared to each other without mentioning it.

From eqs.\ (\ref{eq:BNpp}) and (\ref{eq:GNgg}) it is also clear that any systematic error in the  
$\langle{\cal B}(\psi'\to \chi_c\gamma)\rangle$  
would induce an apparent and technique-correlated
disagreement between the determinations of ${\cal B}(\chi_c\to  p\bar p)$ and
$\Gamma(\chi_c\to \gamma\gamma)$ simultaneously.

\mediumtext
\begin{table}
\begin{tabular}{l c c}
 & \multicolumn{2}{c}{Experimental technique} \\
 & $\psi'\to \chi_c\gamma\to p\bar p\gamma$   
 & $p\bar p\to\chi_c\to J/\psi\gamma$ \\
\hline
${\cal B}(\chi_{c_0}\to p\bar p)$ ($\times10^4$)
          & 1.59$\pm$0.43$\pm$0.53\cite{bib:Bai98I} 
          & 4.8$\pm$0.8$\pm$0.2\cite{bib:Ambrogiani99} \\
${\cal B}(\chi_{c_1}\to p\bar p)$ ($\times10^4$)
             & 0.42$\pm$0.22$\pm$0.28\cite{bib:Bai98I}
             & 0.86$\pm$0.10$\pm$0.11\cite{bib:Armstrong92}\\ 
             & 
             & 0.78$^{+0.18}_{-0.15}\pm0.05$\cite{bib:Baglin86B}\tablenotemark[1]  \\
${\cal B}(\chi_{c_2}\to p\bar p)$ ($\times10^4$)
             & 0.58$\pm$0.31$\pm$0.32\cite{bib:Bai98I}
             & 1.00$\pm$0.08$\pm$0.04\cite{bib:Armstrong92} \\
             & 
             & 0.97$^{+0.44}_{-0.28}\pm$0.08\cite{bib:Baglin86B}  \\
\end{tabular}
\caption{Comparison of ${\cal B}(\chi_c\to p\bar p)$ as obtained by
different experimental techniques}
\label{tab:compp}
\tablenotetext[1]{Calculated as $\Gamma(p\bar p)/\Gamma$ from PDG\cite{bib:PDG00}}
\end{table}

\mediumtext
\begin{table}
\begin{tabular}{l c c c}
& \multicolumn{3}{c}{Experimental technique} \\
&  $\psi'\to \chi_c\gamma\to \gamma\gamma\gamma$ & $p\bar p\to\chi_c\to \gamma\gamma$   
& $\gamma\gamma\to\chi_c\to J/\psi\gamma$\\
\hline
$\Gamma(\chi_{c_0}\to\gamma\gamma)$ (KeV)   
               & 4.0$\pm$2.8\cite{bib:Lee85}
               & 1.45$\pm$0.75$\pm$0.42\cite{bib:Ambrogiani00B}\tablenotemark[1] 
               & \\
$\Gamma(\chi_{c_2}\to\gamma\gamma)$ (KeV)
            & 0.58$\pm$0.31$\pm$0.32\cite{bib:Lee85} 
            & 0.270$\pm$0.049$\pm$0.033\cite{bib:Ambrogiani00B}
            & 1.02$\pm$0.40$\pm$0.17\cite{bib:Acciarri99E} \\
            & 
            & 0.326$\pm$0.08$\pm$0.04\cite{bib:Armstrong93} 
            & 1.76$\pm$0.47$\pm$0.040\cite{bib:Ackerstaff98} \\
            & 
            & 2.0$^{+0.9}_{-0.7}\pm$0.3\cite{bib:Baglin87} 
            & 1.08$\pm$0.30$\pm$0.026\cite{bib:Dominick94} \\
            & 
            & 
            & 3.4$\pm$1.7$\pm$0.9\cite{bib:Bauer93} \\
\end{tabular}
\caption{Comparison of $\Gamma(\chi_c\to\gamma\gamma)$ as obtained by
different experimental techniques}
\label{tab:compg}
\tablenotetext[1]{Calculated using $\Gamma(\chi_{c_0})=14.9^{+2.6}_{-2.4}$ MeV from PDG\cite{bib:PDG00}}
\end{table}

\mediumtext
\begin{table}
\begin{tabular}{l c  l }
Quantity measured & Value & Ref. \\
\hline
$\displaystyle{\frac{{\cal B}(\psi'\to \chi_{c_0}\gamma){\cal B}(\chi_{c_0}\to p\bar p)}
{{\cal B}(\psi'\to J/\psi \pi^+\pi^-)}}$ ($\times 10^{5}$) 
& 4.6$\pm$1.9\tablenotemark[1]\tablenotemark[2] & BES\cite{bib:Bai98I} \\   
\\
$\displaystyle{\frac{{\cal B}(\chi_{c_0}\to \gamma\gamma)}
{{\cal B}(\chi_{c_0}\to J/\psi\gamma)}}$ (\%)
&1.45$\pm$0.74\tablenotemark[2] & E835\cite{bib:Ambrogiani00B}  \\
\\
$\Gamma(\chi_{c_0}\to p\bar p){\cal B}(\chi_{c_0}\to J/\psi\gamma)$ (eV)
& 48.7$\pm$11.5\tablenotemark[2] & E835\cite{bib:Ambrogiani99}  \\
\\
${\cal B}(\psi'\to \chi_{c_0}\gamma){\cal B}(\chi_{c_0}\to J/\psi \gamma)$ (\%) 
& 3.3$\pm$1.7 & SPEAR  \cite{bib:Biddick77} \\
& 0.16$\pm$0.11\tablenotemark[2] & DORIS  \cite{bib:Bartel78B} \\
& 0.4$\pm$0.3\tablenotemark[2]\tablenotemark[4] & DASP   \cite{bib:Brandelik79B} \\
& 0.069$\pm$0.018\tablenotemark[2]\tablenotemark[3] & C.Ball\cite{bib:Oreglia82}   \\
& {\bf 0.073$\pm$0.024} & Average (s=1.3) \\
\\
${\cal B}(\psi'\to \chi_{c_0}\gamma){\cal B}(\chi_{c_0}\to \gamma \gamma)$ ($\times10^5$)
&3.7$\pm$1.8$\pm$1.0 & C.Ball  \cite{bib:Lee85}  \\
\\
${\cal B}(\psi'\to \chi_{c_0}\gamma)$ (\%) 
& 7.5$\pm$2.6 & MARK I \cite{bib:Whitaker76}  \\
& 7.2$\pm$2.3 & SPEAR \cite{bib:Biddick77}  \\
& 9.9$\pm$0.5$\pm$0.8 & C.Ball\cite{bib:Gaiser86}  \\
& {\bf 9.3$\pm$0.8} & Average   \\
\\
$\Gamma(\chi_{c_0})$ (MeV)
& 13.5$\pm$3.3$\pm$4.2  & C.Ball\cite{bib:Gaiser86}  \\
& 14.3$\pm$2.0$\pm$3.0  & BES\cite{bib:Bai98I}  \\
& 16.6$^{+5.2}_{-3.7}\pm$0.1  & E835\cite{bib:Ambrogiani99}  \\
& ${\boldmath{14.9^{+2.6}_{-2.3}}}$  & Average  \\
\end{tabular}
\caption{Available data used in the evaluation of $\chi_{c_0}$ ${\cal B}$'s .}
\tablenotetext[1]{Uses ${\cal B}(\psi' \to J/\psi \pi^+\pi^-)$=0.324$\pm$0.026 
(see \protect\cite{bib:Bai98D}) and 
        ${\cal B}(\psi'\to \chi_{c_0}\gamma)$=0.093$\pm$0.008}
\tablenotetext[2]{Using ${\cal B}(J/\psi\to e^+e^-)$ = 0.0593$\pm$0.0010 and/or 
${\cal B}(J/\psi\to \mu^+\mu^-)$ = 0.0588$\pm$0.0010}
\tablenotetext[3]{Systematic error rescaled for the quoted  ${\cal B}(J/\psi\to l^+l^-)$
contribution and added in quadrature to the statistical error}
\tablenotetext[4]{Uses ${\cal B}(J/\psi\to \mu^+\mu^-)$=0.076$\pm$0.011}
\label{tab:chi0}
\end{table}

\mediumtext
\begin{table}
\begin{tabular}{l c  l }
Quantity measured & Value & Ref. \\
\hline
$\displaystyle{\frac{{\cal B}(\psi'\to \chi_{c_1}\gamma){\cal B}(\chi_{c_1}\to p\bar p)}
{{\cal B}(\psi'\to J/\psi \pi^+\pi^-)}}$ ($\times 10^5$)
& 1.1$\pm$1.0\tablenotemark[1]\tablenotemark[2] & BES\cite{bib:Bai98I} \\   
\\
$\Gamma(\chi_{c_1}\to p\bar p){\cal B}(\chi_{c_1}\to J/\psi\gamma)$ (eV)
& 19.9$\pm$4.4\tablenotemark[2] & R704\cite{bib:Baglin86B}  \\
& 21.7$\pm$2.7\tablenotemark[2] & E760\cite{bib:Armstrong92} \\
& {\bf 21.2$\pm$2.3} & Average  \\
\\
${\cal B}(\psi'\to \chi_{c_1}\gamma){\cal B}(\chi_{c_1}\to J/\psi \gamma)$ (\%)
& 2.8$\pm$0.9\tablenotemark[2] & MARK I \cite{bib:Whitaker76}\cite{bib:Tanenbaum78} \\
& 5.0$\pm$1.5\tablenotemark[5] & SPEAR  \cite{bib:Biddick77} \\
& 2.9$\pm$0.5\tablenotemark[2] & DORIS  \cite{bib:Bartel78B} \\
& 2.2$\pm$0.5\tablenotemark[2]\tablenotemark[6] & DASP   \cite{bib:Brandelik79B}  \\
& 2.78$\pm$0.30\tablenotemark[2]\tablenotemark[3] & C.Ball\cite{bib:Oreglia82} \\
& 2.56$\pm$0.12$\pm$0.20 & C.Ball\cite{bib:Gaiser86}  \\
& {\bf 2.65$\pm$0.16} & Average \\
\\
$\displaystyle{\frac{{\cal B}(\psi'\to \chi_{c_1}\gamma){\cal B}(\chi_{c_1}\to J/\psi\gamma)}
{{\cal B}(\psi'\to J/\psi \pi^+\pi^-)}}$ ($\times 10^2$) 
&8.5$\pm$2.1\tablenotemark[2]\tablenotemark[4]& MARK II\cite{bib:Himel80} \\
\\
${\cal B}(\psi'\to \chi_{c_1}\gamma)$ (\%)
& 7.1$\pm$1.9\tablenotemark[5] & SPEAR \cite{bib:Biddick77} \\
& 9.0$\pm$0.5$\pm$0.7 & C.Ball\cite{bib:Gaiser86}  \\
& {\bf 8.7$\pm$0.8} & Average \\
\\
$\Gamma(\chi_{c_1})$ (MeV)
& 0.88$\pm$0.11$\pm$0.08  & E760\cite{bib:Armstrong92} \\
\end{tabular}
\caption{Available data used in the evaluation of $\chi_{c_1}$ ${\cal B}$'s .}
\tablenotetext[1]{Uses ${\cal B}(\psi' \to J/\psi \pi^+\pi^-)$=0.324$\pm$0.026 (see 
\protect\cite{bib:Bai98D}) and 
        ${\cal B}(\psi'\to \chi_{c_1}\gamma)$=0.087$\pm$0.008}
\tablenotetext[2]{Using ${\cal B}(J/\psi\to e^+e^-)$ = 0.0593$\pm$0.0010 and/or 
${\cal B}(J/\psi\to \mu^+\mu^-)$ = 0.0588$\pm$0.0010}
\tablenotetext[3]{Systematic error rescaled for the quoted  ${\cal B}(J/\psi\to l^+l^-)$
contribution and added in quadrature to the statistical error}
\tablenotetext[4]{Uses ${\cal B}(\psi'\to J/\psi\pi^+\pi^-)$=0.33$\pm$0.03 
(see their ref. [10])}
\tablenotetext[5]{Assumes isotropic $\gamma$ distribution}
\tablenotetext[6]{Uses ${\cal B}(J/\psi\to \mu^+\mu^-)$=0.076$\pm$0.011}
\label{tab:chi1}
\end{table}

\mediumtext
\begin{table}
\begin{tabular}{l c  l }
Quantity measured & Value & Ref. \\
\hline
$\displaystyle{\frac{{\cal B}(\psi'\to \chi_{c_2}\gamma){\cal B}(\chi_{c_2}\to p\bar p)}
{{\cal B}(\psi'\to J/\psi \pi^+\pi^-)}}$ ($\times 10^5$)
&1.4$\pm$1.1\tablenotemark[1]\tablenotemark[2] & BES\cite{bib:Bai98I} \\   
\\
$\displaystyle{\frac{{\cal B}(\chi_{c_2}\to \gamma\gamma)}{{\cal B}(\chi_{c_2}\to J/\psi\gamma)}}$ ($\times10^3$)&
0.99$\pm$0.18\tablenotemark[2] & E835\cite{bib:Ambrogiani00B}  \\
\\
${\cal B}(\chi_{c_2}\to p\bar p){\cal B}(\chi_{c_2}\to \gamma\gamma)$ ($\times 10^8$)
&9.9$\pm$4.5 & R704\cite{bib:Baglin87} \\
&1.60$\pm$0.42 & E760\cite{bib:Armstrong93}  \\
& $\bbox{1.67\pm0.77}$ & Average (s=1.8) \\
\\
$\Gamma(\chi_{c_2}\to p\bar p){\cal B}(\chi_{c_2}\to J/\psi\gamma)$  (eV)
&36$\pm$8\tablenotemark[2] & R704\cite{bib:Baglin86B} \\
&28.2$\pm$2.6\tablenotemark[2] & E760\cite{bib:Armstrong92} \\
&{\bf 28.9$\pm$2.5  } & Average  \\
\\
${\cal B}(\psi'\to \chi_{c_2}\gamma){\cal B}(\chi_{c_2}\to J/\psi \gamma)$  (\%)
& 1.2$\pm$0.7\tablenotemark[2]   & MARK I \cite{bib:Whitaker76} \\
& 2.2$\pm$1.2\tablenotemark[5]   & SPEAR  \cite{bib:Biddick77} \\
& 1.2$\pm$0.2\tablenotemark[2]   & DORIS  \cite{bib:Bartel78B} \\
& 1.8$\pm$0.5\tablenotemark[2]\tablenotemark[6]   & DASP   \cite{bib:Brandelik79B} \\ 
& 1.47$\pm$0.17\tablenotemark[2]\tablenotemark[3] & C.Ball\cite{bib:Oreglia82} \\
& 0.99$\pm$0.10$\pm$0.08 & C.Ball\cite{bib:Gaiser86}  \\
& {\bf 1.20$\pm$0.11} & Average (s=1.2) \\
\\
$\Gamma(\chi_{c_2}\to\gamma\gamma){\cal B}(\chi_{c_2}\to J/\psi \gamma)$ (eV)
&139$\pm$55$\pm$23\tablenotemark[2]\tablenotemark[7]& L3\cite{bib:Acciarri99E}  \\
&240$\pm$64$\pm$55\tablenotemark[2]\tablenotemark[7]& OPAL\cite{bib:Ackerstaff98}  \\
&150$\pm$42$\pm$36\tablenotemark[2]\tablenotemark[7]& CLEO2\cite{bib:Dominick94}  \\
&470$\pm$240$\pm$120\tablenotemark[2]\tablenotemark[7]& TPC$/2\gamma$\cite{bib:Bauer93}  \\
&{\bf 168$\pm$36 } & Average  \\
\\
$\displaystyle{\frac{{\cal B}(\psi'\to \chi_{c_2}\gamma){\cal B}(\chi_{c_2}\to J/\psi\gamma)}
{{\cal B}(\psi'\to J/\psi \pi^+\pi^-)}}$ ($\times 10^2$) 
& 3.9$\pm$1.2\tablenotemark[2]\tablenotemark[4] & MARK II\cite{bib:Himel80}   \\
\\
${\cal B}(\psi'\to \chi_{c_2}\gamma){\cal B}(\chi_{c_2}\to \gamma \gamma)$ ($\times 10^5$)
&7.0$\pm$2.1$\pm$2.0 & C.Ball  \cite{bib:Lee85}  \\
\\
${\cal B}(\psi'\to \chi_{c_2}\gamma)$ (\%) 
& 7.0$\pm$2.0\tablenotemark[5] & SPEAR \cite{bib:Biddick77}  \\
& 8.0$\pm$0.5$\pm$0.7 & C.Ball\cite{bib:Gaiser86}  \\
& {\bf 7.8$\pm$0.8}          & Average  \\
\\
$\Gamma(\chi_{c_2})$ (MeV)
& 1.98$\pm$0.17$\pm$0.07 & E760\cite{bib:Armstrong92}  \\
& 2.6$^{+1.4}_{-1.0}$  & R704\cite{bib:Baglin86B}  \\
& 2.4$^{+2.1}_{-2.0}$  & C.Ball\cite{bib:Gaiser86}  \\
& {\bf 2.0$\pm$0.18}   & Average \\
\end{tabular}
\caption{Available data used in the evaluation of $\chi_{c_2}$ ${\cal B}$'s .}
\tablenotetext[1]{Uses ${\cal B}(\psi' \to J/\psi \pi^+\pi^-)$=0.324$\pm$0.026 
(see \protect\cite{bib:Bai98D}) and 
        ${\cal B}(\psi'\to \chi_{c_2}\gamma)$=0.078$\pm$0.008}
\tablenotetext[2]{Using ${\cal B}(J/\psi\to e^+e^-)$ = 0.0593$\pm$0.0010 and/or 
${\cal B}(J/\psi\to \mu^+\mu^-)$ = 0.0588$\pm$0.0010}
\tablenotetext[3]{Systematic error rescaled for the quoted  ${\cal B}(J/\psi\to l^+l^-)$
contribution and added in quadrature to the statistical error}
\tablenotetext[4]{Uses ${\cal B}(\psi'\to J/\psi\pi^+\pi^-)$=0.33$\pm$0.03 
(see their ref. [10])} 
\tablenotetext[5]{Assumes isotropic $\gamma$ distribution}
\tablenotetext[6]{Uses ${\cal B}(J/\psi\to \mu^+\mu^-)$=0.076$\pm$0.011}
\tablenotetext[7]{Uses ${\cal B}(\chi_{c_2}\to J/\psi\gamma)$=0.135$\pm$0.011}
\label{tab:chi2}
\end{table}

\mediumtext
\begin{table}
\begin{tabular}{l c  l }
Quantity measured & Value & Ref. \\
\hline
${\cal B}(\psi'\to J/\psi\, X)$ 
& 0.51$\pm$0.12                & DASP \cite{bib:Brandelik79C}  \\
& 0.57$\pm$0.08                & MARK I \cite{bib:Abrams75B}  \\
&{\bf 0.55$\pm$0.07}           & Average  \\ 
\\
${\cal B}(\psi'\to J/\psi \pi^+\pi^-)$ & 0.32$\pm$0.04    &  MARK I  
\cite{bib:Abrams75B} \\
\\
${\cal B}(\psi'\to J/\psi \eta)$ 
& 0.042$\pm$0.006\tablenotemark[1] & DORIS \cite{bib:Bartel78B}  \\
& 0.045$\pm$0.012\tablenotemark[1]\tablenotemark[2] & DASP \cite{bib:Brandelik79B} \\ 
& 0.0255$\pm$0.0029\tablenotemark[1]\tablenotemark[3] & C.Ball
\cite{bib:Oreglia80}\cite{bib:Oreglia82}  
 \\
& {\bf 0.0294$\pm$0.0051 }& Average (s=2.0) \\
\\
${\cal B}(\psi'\to \chi_{c_1}\gamma\to J/\psi \gamma\gamma)$ 
& 0.028$\pm$0.009\tablenotemark[1]  & MARK I \cite{bib:Whitaker76}\cite{bib:Tanenbaum78} \\
& 0.050$\pm$0.015\tablenotemark[4]  & SPEAR  \cite{bib:Biddick77} \\
& 0.029$\pm$0.005\tablenotemark[1]  & DORIS  \cite{bib:Bartel78B} \\
& 0.022$\pm$0.005\tablenotemark[1]\tablenotemark[2] & DASP   \cite{bib:Brandelik79B} \\
& 0.0278$\pm$0.0030\tablenotemark[1]\tablenotemark[3] & C.Ball\cite{bib:Oreglia82} \\
& 0.0256$\pm$0.0012$\pm$0.0020                    & C.Ball\cite{bib:Gaiser86} \\
& {\bf 0.0265$\pm$0.0016} & Average \\
\\
${\cal B}(\psi'\to \chi_{c_2}\gamma\to J/\psi \gamma\gamma)$ 
& 0.012$\pm$0.007\tablenotemark[1]    & MARK I \cite{bib:Whitaker76} \\
& 0.022$\pm$0.012\tablenotemark[4]    & SPEAR  \cite{bib:Biddick77}  \\
& 0.012$\pm$0.002\tablenotemark[1]    & DORIS  \cite{bib:Bartel78B} \\
& 0.018$\pm$0.005\tablenotemark[1]\tablenotemark[2] & DASP  \cite{bib:Brandelik79B} \\ 
& 0.0147$\pm$0.0017\tablenotemark[1]\tablenotemark[3] & C.Ball\cite{bib:Oreglia82} \\
& 0.0099$\pm$0.0010$\pm$0.0008                   & C.Ball\cite{bib:Gaiser86} \\
& {\bf 0.0120$\pm$0.0011} & Average (s=1.2) \\
\\
$\displaystyle{\frac{{\cal B}(\psi'\to J/\psi \pi^+\pi^-)}{{\cal B}(\psi'\to J/\psi\, X)}}$ 
& 0.496$\pm$0.037\tablenotemark[6]  & E-760 
\cite{bib:Armstrong97} \\
\\
$\displaystyle{\frac{{\cal B}(\psi'\to J/\psi\, neutrals)}{{\cal B}(\psi'\to J/\psi \pi^+\pi^-)}}$  
& 0.73$\pm$0.09    & MARK I  
\cite{bib:Tanenbaum76}   \\
\\
$\displaystyle{\frac{{\cal B}(\psi'\to J/\psi \pi^0\pi^0)}{{\cal B}(\psi'\to J/\psi\, X)}}$ 
& 0.323$\pm$0.033\tablenotemark[6]  & E-760 \cite{bib:Armstrong97}  \\
& 0.328$\pm$0.015  & E-835 \cite{bib:Ambrogiani00} \\
& {\bf 0.327$\pm$0.014}  & Average  \\
\\
$\displaystyle{\frac{{\cal B}(\psi'\to J/\psi \eta)}{{\cal B}(\psi'\to J/\psi \pi^+\pi^-)}}$ 
& 0.091$\pm$0.021\tablenotemark[1]\tablenotemark[5]  & MARK II\cite{bib:Himel80}  \\
\\
$\displaystyle{\frac{{\cal B}(\psi'\to J/\psi \eta)}{{\cal B}(\psi'\to J/\psi\, X)}}$ 
& 0.061$\pm$0.015\tablenotemark[6]  & E-760\cite{bib:Armstrong97}  \\
& 0.072$\pm$0.009  & E-835\cite{bib:Ambrogiani00} \\
& {\bf 0.069$\pm$0.008 } &Average \\
\\
$\displaystyle{\frac{{\cal B}(\psi'\to \chi_{c_1}\gamma\to J/\psi\gamma\gamma)}
{{\cal B}(\psi'\to J/\psi \pi^+\pi^-)}}$ 
& 0.085$\pm$0.021\tablenotemark[1]\tablenotemark[5]  & MARK II\cite{bib:Himel80} \\
\\
$\displaystyle{\frac{{\cal B}(\psi'\to \chi_{c_2}\gamma\to J/\psi\gamma\gamma)}
{{\cal B}(\psi'\to J/\psi \pi^+\pi^-)}}$ 
& 0.039$\pm$0.012\tablenotemark[1]\tablenotemark[5]   & MARK II\cite{bib:Himel80} \\
\end{tabular}
\caption{Available data used in the evaluation of $\psi'$ ${\cal B}$'s}
\tablenotetext[1]{Using ${\cal B}(J/\psi\to e^+e^-)$ = 0.0593$\pm$0.0010 and/or 
${\cal B}(J/\psi\to \mu^+\mu^-)$ = 0.0588$\pm$0.0010}
\tablenotetext[2]{Uses ${\cal B}(J/\psi\to \mu^+\mu^-)$=0.076$\pm$0.011}
\tablenotetext[3]{Systematic error rescaled for the quoted ${\cal B}(J/\psi\to l^+l^-)$ contribution and added in quadrature
to the statistical error.}
\tablenotetext[4]{Assumes isotropic $\gamma$ distribution}
\tablenotetext[5]{Uses ${\cal B}(\psi'\to J/\psi\pi^+\pi^-)$=0.33$\pm$0.03 
(see their ref. [10])}
\tablenotetext[6]{Calculated from efficiency and event ratios}
\label{tab:measp}
\end{table}

\widetext
\begin{table}
\begin{tabular}{l c c c c c }
Parameter & ${\cal B}(\psi'\to \chi\gamma)$ 
                       & ${\cal B}(\psi'\to J/\psi\pi^+\pi^-)$ 
                                             &$\Gamma$ fixed & New fit & PDG \cite{bib:PDG00} \\
           & fixed         & fixed           &               &         &                      \\
           &               & (separate fits) &               &         &                      \\
\hline
$\Gamma(\chi_{c_0})$ (MeV)  & 16.3$\pm$2.3 & 15.5$\pm$2.4 & 14.9 (fixed)  &  15.5$\pm$2.4  & 14.9$^{+2.6}_{-2.3}$   \\
$\Gamma(\chi_{c_1})$ (MeV) & 0.92$\pm$0.13 & 0.91$\pm$0.13 & 0.88 (fixed)  &  0.92$\pm$0.13 & 0.88$\pm$0.11$\pm$0.07  \\
$\Gamma(\chi_{c_2})$ (MeV) & 2.07$\pm$0.17 & 2.07$\pm$0.17 & 2.00 (fixed)  &  2.07$\pm$0.17 & 2.0$\pm$0.18  \\
\\
${\cal B}(\psi'\to J/\psi \pi^+\pi^-)$ (\%)            
& 30.8$\pm$2.2 & 32.0 (fixed) & 30.3$\pm$2.2   & 30.2$\pm$2.2  & 31.0$\pm$2.8\\
${\cal B}(\psi'\to J/\psi \pi^0\pi^0)$ (\%)            
&18.3$\pm$1.5 &- & 18.1$\pm$1.5   & 18.0$\pm$1.5  & 18.2$\pm$2.3 \\
${\cal B}(\psi'\to J/\psi \eta)$      (\%)                  
&3.14$\pm$2.2 & - & 3.12$\pm$0.22  & 3.10$\pm$0.22 & 2.7$\pm$0.4 \\
${\cal B}(\psi'\to \chi_{c_0}\gamma)$  (\%) 
&9.3 (fixed) & 7.2$\pm$1.2 & 7.1$\pm$1.1 & 7.1$\pm$1.2 & 9.3$\pm$0.9  \\
${\cal B}(\psi'\to \chi_{c_1}\gamma)$  (\%) 
&8.7 (fixed) & 8.4$\pm$0.8 & 8.4$\pm$0.8 & 8.4$\pm$0.8  & 8.7$\pm$0.8   \\
${\cal B}(\psi'\to \chi_{c_2}\gamma)$  (\%) 
&7.8 (fixed) & 6.7$\pm$0.8 & 6.7$\pm$0.8 & 6.8$\pm$0.8  & 7.8$\pm$0.8    \\
${\cal B}(\psi'\to J/\psi\, X$  (\%)\tablenotemark[1] 
&56$\pm$4 & - & 55$\pm$5 & 55$\pm$5  &  55$\pm$5   \\
\\
${\cal B}(\chi_{c_0}\to J/\psi\gamma)$ (\%) 
&1.0$\pm$0.2$\pm$0.1\tablenotemark[2] & 1.1$^{+0.3}_{-0.2}$ & 1.2$^{+0.3}_{-0.2}$ &1.2$^{+0.3}_{-0.2}$ & 
0.66$\pm$0.18$\pm$0.06\tablenotemark[2] \\
${\cal B}(\chi_{c_0}\to p\bar p)$ ($\times 10^4$) 
&2.2$\pm$0.5$\pm$0.2\tablenotemark[2] & 2.5$\pm$0.6 &2.5$\pm$0.6 & 2.4$\pm$0.6 &
2.2$\pm$1.3$\pm$0.2\tablenotemark[2]   \\ 
${\cal B}(\chi_{c_0}\to \gamma\gamma)$ ($\times 10^4$) 
&1.6$\pm$0.7$\pm$0.2\tablenotemark[2] & 2.0$^{+1.0}_{-0.9}$ & 2.0$^{+1.0}_{-0.9}$ & 2.0$^{+1.0}_{-0.9}$ &  
2.7$\pm$1.9$\pm$0.3\tablenotemark[2]  
\\ 
\\
${\cal B}(\chi_{c_1}\to J/\psi\gamma)$ (\%) 
& 30.7$\pm$1.7$\pm$2.8\tablenotemark[2] &31.8$^{+3.6}_{-3.1}$ & 31.9$^{+3.6}_{-3.1}$ &31.8$^{+3.6}_{-3.1}$ &
27.3$\pm$1.6$\pm$2.5\tablenotemark[2] \\
${\cal B}(\chi_{c_1}\to p\bar p)$ ($\times 10^5$) 
&7.3$^{+1.4}_{-1.2}\pm$0.7\tablenotemark[2] & 7.1$^{+1.5}_{-1.2}$ &7.4$^{+1.1}_{-1.0}$ & 7.1$^{+1.5}_{-1.2}$ &  
8.2$\pm$1.3$\pm$0.8\tablenotemark[2]  \\ 
\\
${\cal B}(\chi_{c_2}\to J/\psi\gamma)$ (\%) 
&16.3$\pm$1.0$\pm$1.6\tablenotemark[2] &18.7$^{+2.8}_{-2.3}$& 18.8$^{+2.8}_{-2.4}$ &18.7$^{+2.8}_{-2.3}$ & 
13.5$\pm$1.1$\pm$1.4\tablenotemark[2] \\
${\cal B}(\chi_{c_2}\to p\bar p)$ ($\times 10^5$) 
&8.5$^{+1.2}_{-1.0}\pm$0.9\tablenotemark[2] & 7.5$^{+1.4}_{-1.2}$ &7.7$^{+1.3}_{-1.1}$ & 7.4$^{+1.4}_{-1.2}$ &
9.8$\pm$1.0$\pm$1.0\tablenotemark[2]   \\ 
${\cal B}(\chi_{c_2}\to \gamma\gamma)$ ($\times 10^4$) 
&2.0$\pm$0.3$\pm$0.2\tablenotemark[2] & 2.2$\pm$0.4 &2.2$\pm$0.4 & 2.2$\pm$0.4& 
1.6$\pm$0.5$\pm$0.2\tablenotemark[2]   \\
\\
$\chi^2/N_{DOF}$    
&58.6/42 & 9.1/9  ($\chi_{c_0}$) & 55.9/42   & 56/46  &   \\
& & 5.7/9  ($\chi_{c_1}$) & &   &    \\
& & 26.5/18 ($\chi_{c_2}$) & &   &    \\

\end{tabular}
\caption{$\psi'$ and $\chi_{c}$ ${\cal B}$'s as determined from the fits described in the text.
The second column gives the results which would be obtained by ignoring the correlation to
${\cal B}(\psi'\to J/\psi \pi^+\pi^-)$ and performing a separate fit for each one of the $\chi_c$ 
states}
\tablenotetext[1]{Calculated as 
${\cal B}(\psi'\to J/\psi\pi^+\pi^-)+
 {\cal B}(\psi'\to J/\psi\pi^0\pi^0)  + 
 {\cal B}(\psi'\to J/\psi\eta) +
 {\cal B}(\psi'\to\chi_{c_1}\gamma){\cal B}(\chi_{c_1}\to J/\psi\gamma)+ 
 {\cal B}(\psi'\to\chi_{c_2}\gamma){\cal B}(\chi_{c_2}\to J/\psi\gamma)$ 
}
\tablenotetext[2]{The second error refers to the uncertainty in the 
${\cal B}(\psi'\to\chi_c\gamma)$}
\label{tab:fglob}
\end{table}

\narrowtext
\begin{table}
\begin{tabular}{l c c c }
Transition                  & \multicolumn{2}{c}{Experiment}   & Predicted \\
                            & PDG & This fit         &  \\
\hline
$\psi'\to\chi_{c_0}\gamma$  & 26$\pm$4   & 19$\pm$4         & 21\cite{bib:Fayya} \\
                            &            &                  & 19.8\cite{bib:Gupta89} \\
                            &            &                  & 31(32)\cite{bib:Resag95}\\
                            &            &                  & 19(16)\cite{bib:McClary}\\
$\psi'\to\chi_{c_1}\gamma$  & 24$\pm$3   & 23$\pm$4         & 20\cite{bib:Fayya} \\
                            &            &                  & 29.4\cite{bib:Gupta89} \\
                            &            &                  & 36(48)\cite{bib:Resag95}\\
                            &            &                  & 31(23)\cite{bib:McClary}\\
$\psi'\to\chi_{c_2}\gamma$  & 21$\pm$3   & 19$\pm$3         & 14\cite{bib:Fayya} \\
                            &            &                  & 25.4\cite{bib:Gupta89} \\
                            &            &                  & 60(35)\cite{bib:Resag95}\\
                            &            &                  & 27(22)\cite{bib:McClary}\\
\\
$\chi_{c_0}\to J/\psi\gamma$& 98$\pm$34  & 179$\pm$55       & 141\cite{bib:Fayya} \\
                            & 117$\pm$45\tablenotemark[1] & & 116.6 \cite{bib:Gupta89} \\
                            &            &                  & 140(119)\cite{bib:Resag95}\\
                            &            &                  & 128(117)\cite{bib:McClary}\\
$\chi_{c_1}\to J/\psi\gamma$& 240$\pm$44 & 282$\pm$52       & 273\cite{bib:Fayya} \\
                            & 268$\pm$49\tablenotemark[1] & & 249.5\cite{bib:Gupta89} \\
                            &            &                  & 250(230)\cite{bib:Resag95}\\
                            &            &                  & 270(240)\cite{bib:McClary}\\
$\chi_{c_2}\to J/\psi\gamma$& 270$\pm$43 & 374$\pm$62       & 347\cite{bib:Fayya} \\
                            & 306$\pm$48\tablenotemark[1] & & 326.1\cite{bib:Gupta89} \\
                            &            &                  & 270(347)\cite{bib:Resag95}\\
                            &            &                  & 347(305)\cite{bib:McClary}\\
\\
$\chi_{c_0}\to \gamma\gamma$& 1.9$\pm$1.0\tablenotemark[2]   & 3.0$\pm$1.5    & 3.72$\pm$0.11\cite{bib:Huang} \\
                            &                                &                & 6.38\cite{bib:Gupta} \\
                            &                                &                & 1.39$\pm$0.16\cite{bib:Muntz} \\
                            &                                &                & 11.3$^{+4.7}_{-4.0}$\cite{bib:Bodwin} \\
$\chi_{c_2}\to \gamma\gamma$& 0.32$\pm$0.09\tablenotemark[2] & 0.44$\pm$0.09  & 0.49$\pm$0.15\cite{bib:Huang}\\
                            &                                &                & 0.57\cite{bib:Gupta} \\
                            &                                &                & 0.44$\pm$0.14\cite{bib:Muntz} \\
                            &                                &                & 0.83$\pm$0.23\cite{bib:Bodwin} \\
\end{tabular}
\caption{Comparison of partial widths with some theoretical expectations. All values are in KeV.}
\label{tab:Theo}
\tablenotetext[1]{Calculated using the $\chi_c$ branching ratios wich would be obtained  
following the PDG procedure with the new value of ${\cal B}(J/\psi\to l^+l^-)$.}
\tablenotetext[2]{Using also E835\cite{bib:Ambrogiani00B} result in the average}
\end{table}

\end{document}